\documentclass[%
reprint,
superscriptaddress,
frontmatterverbose,
showpacs,preprintnumbers,
 amsmath,amssymb,
 aps,prl,Ifloatfix,
citeautoscript,
]{revtex4-2}

\usepackage[pdftex]{graphicx}
\usepackage{mathrsfs}
\usepackage[linkcolor=blue,colorlinks=true,citecolor=blue,filecolor=blue]{hyperref}
\usepackage{amsmath}
\usepackage[english]{babel}
\usepackage{booktabs}
\usepackage{amssymb}
\usepackage{type1cm}
\usepackage{braket}
\usepackage{tikz}
\usepackage{pgfplots}
\usepackage{lipsum}
\usepackage{xcolor}
\usepackage{graphicx}% Include figure files
\usepackage{subcaption}
\usepackage{dcolumn}% Align table columns on decimal point
\usepackage{bm}% bold mathd
\usepackage{physics}
\usepackage{float}
\usepackage[utf8]{inputenc}
\usepackage{multirow}

\definecolor{Midnight_Blue}{rgb}{0.1, 0.1, 0.6}

\usepackage{cleveref}
\crefname{equation}{Eq.}{Eqs.}
\Crefname{equation}{Equation}{Equations}
\crefname{figure}{Fig.}{Figs.}
\Crefname{figure}{Figure}{Figures}
\crefname{figure}{Fig.}{Figs.}
\Crefname{figure}{Figure}{Figures}
\crefname{section}{Supplemental Material Section}{Supplemental Material Sections}
\Crefname{section}{Supplemental Material Section}{Supplemental Material Sections}
\crefname{appendix}{Appendix}{Appendices}
\Crefname{appendix}{Appendix}{Appendices}
\crefname{table}{Table}{Tables}
\Crefname{table}{Table}{Tables}

\usepackage{enumitem,amssymb}
\newlist{todolist}{itemize}{2}
\setlist[todolist]{label=$\square$}
\usepackage{pifont}

\usepackage{etoolbox}

\begin{document}

\title{Improving Continuous-variable Quantum Channels with Unitary Averaging}
%\title{Universal Quantum Information through Passive Unitary Averaging with Continuous-variable squeezed light}

\author{S. Nibedita Swain}
\affiliation{School of Mathematical and Physical Sciences,
University of Technology Sydney, Ultimo, NSW 2007, Australia}

\affiliation{Sydney Quantum Academy, Sydney, NSW 2000, Australia}
\affiliation{Centre for Quantum Computation and Communication Technology, School of Mathematics
and Physics, University of Queensland, Brisbane, Queensland 4072, Australia}
\author{Ryan J. Marshman}
\affiliation{Centre for Quantum Computation and Communication Technology, School of Mathematics
and Physics, University of Queensland, Brisbane, Queensland 4072, Australia}
\author{Peter P. Rohde}
\affiliation{ BTQ \& Centre for Engineered Quantum Systems, Macquarie University, Sydney NSW, Australia}
\author{Austin P. Lund}
\affiliation{Centre for Quantum Computation and Communication Technology, School of Mathematics
and Physics, University of Queensland, Brisbane, Queensland 4072, Australia}
\affiliation{Xanadu, Toronto, Ontario, M5G 2C8, Canada}
\author{Alexander S. Solntsev}
\affiliation{School of Mathematical and Physical Sciences,
University of Technology Sydney, Ultimo, NSW 2007, Australia}

\author{Timothy C. Ralph}
\affiliation{Centre for Quantum Computation and Communication Technology, School of Mathematics
and Physics, University of Queensland, Brisbane, Queensland 4072, Australia}

\begin{abstract}
A significant hurdle for quantum information and processing using bosonic systems is stochastic phase errors which occur as the photons propagate through a channel. These errors will reduce the purity of states passing through the channel and so reducing the channels capacity. We present a scheme of passive linear optical unitary averaging for protecting unknown Gaussian states transmitted through an optical channel. The scheme reduces the effect of phase noise on purity, squeezing and entanglement, thereby enhancing the channel via probabilistic error correcting protocol. The scheme is robust to loss and typically succeeds with high probability. We provide both numerical simulations and analytical approximations tailored for relevant parameters with the improvement of practical and current technology. We also show the asymptotic nature of the protocol, highlighting both current and future relevance.

\end{abstract}

\date{\today}

\frenchspacing

\maketitle

%\section{Introduction}

\textit{Introduction.-} Optical quantum systems play a major role in a wide range of quantum technology applications %computing \cite{bourassa2021blueprint} and communication architectures \cite{weedbrook2012gaussian, ferraro2005gaussian}, 
offering distinct advantages \cite{BAC19}. Currently, some of the most developed and practical quantum information applications of optical systems are based on Gaussian source states \cite{weedbrook2012gaussian} (such as squeezed states) and linear networks. Examples include continuous variable (CV) versions of teleportation \cite{FUR98} and quantum key distribution \cite{GRO03} as well as Gaussian Boson sampling \cite{lund2014boson,Hamilton2017_GBS,ZHO20,MAD22}.
%They are particularly valuable for simulating bosonic systems (e.g., boson sampling- \cite{aaronson2011computational, lund2014boson,lund2017quantum}) and modeling the quantum vibrational dynamics of molecules \cite{sparrow2018simulating, mcardle2020quantum}.
% The potential of this technology is warranted by its ability to function at room temperature, ease of manufacturability, resilience to photon loss, and also for scalability \cite{kok2007linear}. 
However, in order to enable scalable quantum applications it is essential to establish practical approaches for controlling noise in these quantum optical systems. %Quantum computing designs and error-correction methods that can minimise noise would be highly beneficial in advancing this pursuit \cite{tzitrin2021fault}.
The non-universal nature of deterministic linear optics processing means standard approaches to error correction are not immediately applicable and other approaches need to be explored.

One such alternative approach is unitary averaging (UA) \cite{marshman2018passive, vijayan2020robust, marshman2023unitary, singh2022optical}, a framework that has been shown to help reduce errors within discrete variable (DV) linear optical setups, i.e. set-ups in which non-Gaussian, single photon source states are injected into linear networks. It is not immediately obvious that UA can be usefully extended to systems with Gaussian inputs due to various Gaussian no-go theorems \cite{fiuravsek2002gaussian, giedke2002characterization,niset2009no}. The role of vacuum projection with respect to the theorems is non-trivial.
None-the-less we show that such an extension is possible and useful, leading to a powerful generalization of the UA technique. %In extending the passive averaging \cite{marshman2018passive} from discrete to continuous variable systems, we show that passive averaging, which is an unitary in this context, acts as a noise filter against phase errors. Furthermore, we also present results considering loss.

To illustrate the effect we focus on a simple but practically relevant example: a single mode propagating through a channel with stochastic phase noise. Such a situation is generic in continuous variable quantum communication scenarios but might also be relevant in optical circuits due to fabrication variations or the incorporation of thermal components. We  characterise the improvement in the channel by analysing a two-mode squeezed vacuum state \cite{PhysRevLett.68.3663} with half the state sent through the channel. Previous techniques have employed multiple copies of the input state to reduce phase noise \cite{franzen2006experimental, fiuravsek2007experimentally, lassen2010experimental}. In contrast, here a single copy of an unknown state is sent through multiple copies of the channel before being recombined non-deterministically. We demonstrate that Gaussian encoding with vacuum projection (as used for recombination) has strong utility for protecting Gaussian states against phase errors. This leads to a significant improvement of purity, entanglement and squeezing through the averaged channel with a high probability of success. Importantly, the protocol continues to perform effectively in the presence of loss in the optical elements.

%\section{Unitary averaging on DV systems}
\textit{Unitary averaging on DV systems.-} In the UA framework, one utilises redundancy in the applied transformation, rather than redundant encoding to protect against errors. The success probability is known to depend on the variance in the individual unitaries.
UA acts to apply an averaged unitary evolution \cite{marshman2018passive,vijayan2020robust} on discrete variable (DV) systems given by
\begin{align}
    \mathnormal{\hat{u}} = \frac{1}{N}\sum^{N}_{k =1} \hat{U}_{k}
\end{align}
using $N$ copies of the unitary. The choice of $\mathnormal{\hat{u}}$ over $\hat{U}$ signifies that the resulting transformation in DV systems is non-unitary. Additionally, there are a set of $(N-1)$ `error' modes that are post-selected by vacuum projections to herald the success of this transformation. This average transformation $\mathnormal{\hat{u}}$  represents a stochastic operator which approximates the target unitary with a variance reduced by a factor $N$ \cite{marshman2018passive}. For DV optics this was found to allow a trade-off between the transformation fidelity and the the heralded probability of success.

%The unnormalised output state to only considering the correct modes is
%\begin{align}
%    \hat{\rho}(N) = \mathnormal{\hat{u}}(N)\ket{\psi}\bra{\psi}\mathnormal{\hat{u}}^{\dagger}(N)
%\end{align}
%with the success probability given by the magnitude of the unnormalised state
%which is given by
%\begin{align}
  %  P_{s}(N) & = | \mathnormal{\hat{u}}(N)\ket{\psi}|^{2} \nonumber\\
    %& = \frac{1}{N^{2}} \sum^{N}_{k =1}\sum^{N}_{l=1}\langle \psi |  \hat{U}^{\dagger}_{k} \hat{U}_{l}|\psi \rangle
%\end{align}
%Another important aspect for assessing the impact of UA on DV system is the state fidelity. It is defined as,
%\begin{align}
  %  \mathcal{F}(N) = \bra{\psi} \hat{\rho}(N)\ket{\psi}
%\end{align}
% We will now extend the unitary averaging model to continuous variable systems and 
We will now apply UA to a single mode channel suffering stochastic phase noise.

\begin{figure}
    \includegraphics[scale=0.4]{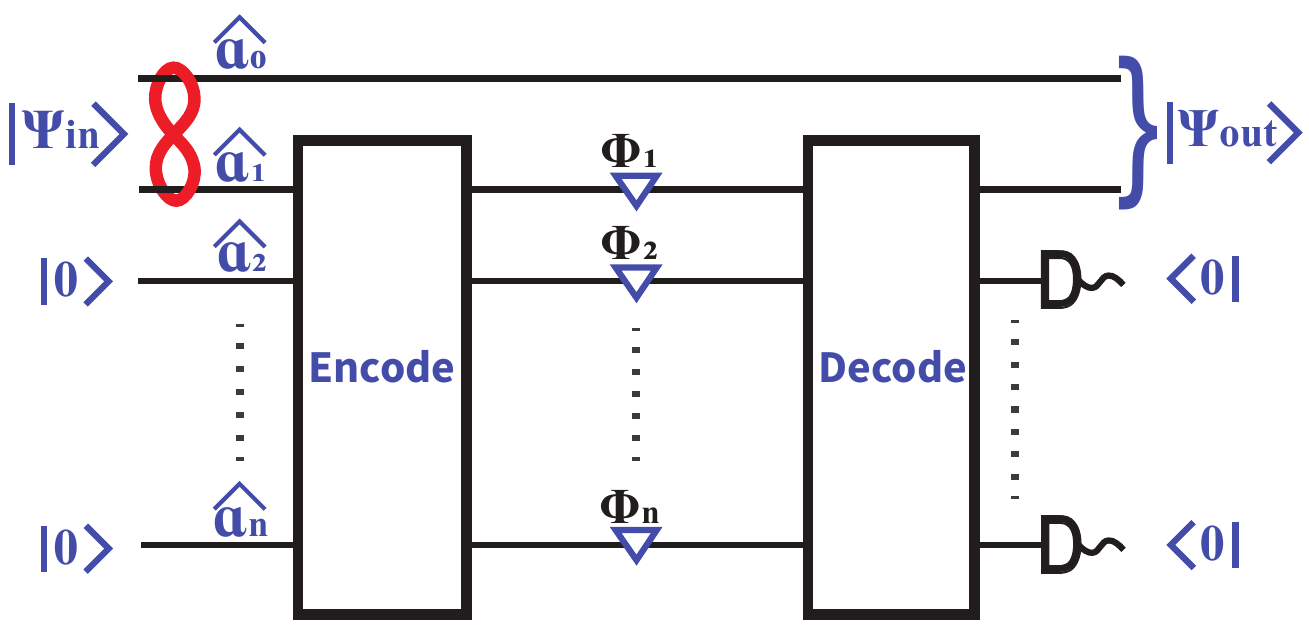}
    \caption{Scheme for passive unitary averaging. Redundant encoding using beamsplitter network for general $n$. The encoding beamsplitter and phase shifter network are repeated for each mode. Then post-selected on the vacuum in the (n-1)th mode. Specifically, we apply a single mode $(\hat{a}_{1})$ of two-mode squeezed vacuum state as the input state in mode-1, the other mode is free $(\hat{a}_{0})$ and post-selected on (n-1)th mode to get the output state.}
    \label{cir}
\end{figure}

%\section{Passive unitary averaging on CV systems}
 \textit{Passive unitary averaging on CV systems.-} We will model the effect of UA on a CV system by considering a two-mode squeezed vacuum state \cite{schumaker1985new, lvovsky2015squeezed}, with one mode fed through the noisy channel which is then characterised by the output squeezing, purity and entanglement. The two mode squeezed vacuum was chosen as it can be used to derive a general expression for many other states through projective measurement. Fig.\ref{cir} shows the circuit we will be considering. We direct one mode of the two-mode squeezed vacuum state into the interferometer, while leaving the other mode free. The encode and decode transformations consist of $50:50$ beamsplitters implementing a Hadamard transform to evenly mix all modes together. Finally, the $(n-1)$ output modes are heralded in the vacuum leaving the signal to be output through the final, undetected mode.
 %We will characterise the effect of UA on a channel by considering a two-mode squeezed vacuum states \cite{schumaker1985new, lvovsky2015squeezed}, with one mode fed through the system and the resulting squeezing, purity and entanglement used to characterise the channel. The two mode squeezed vacuum was chosen as it can be used to derive a general expression for many other states through projective measurement. Fig.\ref{cir} shows the circuit we will be considering. We direct one mode of the two-mode squeezed vacuum state into the interferometer, while leaving the other mode free. The encode and decode transformations consist of $50:50$ beamsplitters implementing a Hadamard transform to evenly mix all modes together. Finally, the $(n-1)$ output modes are heralded in the vacuum leaving the signal to be output through the final, undetected mode.
% The model under consideration assumes that large linear networks can be arbitrarily configured, albeit with some introduced Gaussian phase noise.

The input state is given by
\begin{align}
      \ket{\psi_{\text{in}}} = \hat{S}(r)\ket{0}\ket{0} \otimes \ket{0}^{\otimes\left(n-1\right)} \label{inputeq}
\end{align}
where the two-mode squeezing operator $\hat{S}(r)$ acts on only the first two modes %with $\chi=re^{i\phi}$ and
 \begin{align}
  \hat{S}(r) = \exp\left[-r(\hat{a}_{0}\hat{a}_{1} -  \hat{a}_{0}^{\dag}\hat{a}_{1}^{\dag}) \right]   
\end{align}
After passing through the interferometer and heralding on the last $(n-1)$ modes, the output state, Eq.\ref{inputeq} is
 \begin{align}
    \left(\otimes_{j=2}^n\bra{0}_j\right)\hat{U} \hat{S}(r)\ket{0}\ket{0} \otimes \ket{0}^{\otimes(n-1)} = \ket{\psi_{\text{out}}}
 \end{align}
 where $\hat{U} = (\hat{H}^{\dagger}*\hat{R}(\theta)*\hat{H})_{n \times n}$,  $\hat{H}$ is the Hadamard transforms acting as the encoding and decoding circuits, and $\hat{R}(\theta)$ implements stochastic phase transformations on each mode, representing the noise.
 The output state is,
 \begin{widetext}
     \begin{equation}
     \ket{\psi_{\text{out}}} = (\cosh{r})^{-1} \sum_{N} (-1)^{j}\left[\frac{e^{ i \phi_{1}} + e^{ i \phi_{2}} +...+e^{ i \phi_{n}}}{n} \tanh{r}\right]^{N} \ket{N, N} \label{outputeq}
 \end{equation}

\end{widetext}

We will simplify this by defining
\begin{align}
\alpha e^{i\phi_{\beta}}=&\frac{e^{ i \phi_{1}} + e^{ i \phi_{2}} +...+e^{ i \phi_{n}}}{n}\\
    \tanh{r'} =& \alpha\tanh{r}
\end{align}
The phase terms $\phi_j$ vary randomly and independently around some mean value.
Hence,  the  output state after post selection becomes
\begin{align}
    \ket{\psi_{out}} &= \frac{1}{\mathbf{N}} \left(\cosh{r}\right)^{-1}\sum_{N} (-1)^{N} [e^{i \phi_{\beta}} \tanh{r'} ]^{N} \ket{N, N}\nonumber\\
    &=\hat{S}\left(\chi'\right)\ket{0}\ket{0}
\end{align}
where $\mathbf{N}$ is a normalisation constant related to probability of success, and $\chi'=r'e^{i\phi_{\beta}}$.

% Now, we present an analytical formula,
% \begin{align}
%     \hat{S}\left(\chi'\right) = \cosh{(2r')} -\sinh{(2r')}\cos{(2\sqrt{\frac{v}{n}})}
% \end{align}
The normalisation constant for the output state $(\ref{outputeq})$ is,
\begin{align}
    \mathbf{N} 
    & = \frac{\cosh{r'}}{\cosh{r}}
\end{align}
giving a probability of success  $P = |\frac{\cosh{r'}}{\cosh{r}}|^{2}$.

While each individual use of the channel will produce two mode squeezed state with squeezing parametrised by $\chi'$, the values of $\alpha$, $\phi_{\beta}$ and hence $\chi'$ are each stochastic variables. As such, the output is the mixed state 
\begin{equation}
    \hat{\rho}=\int \mathrm{d}\chi' p(\chi') \hat{S}\left(\chi'\right)\ket{0}\ket{0}\bra{0}\bra{0}\hat{S}^{\dagger}\left(\chi'\right)
\end{equation}
where $p(\chi')$ gives the probability density for the stochastic parameter $\chi'$. Thus, the state is only approximately Gaussian for small noise, with the approximation becoming exact in the limit of no noise. 

The term \textit{Gaussian} \cite{weedbrook2012gaussian, ferraro2005gaussian} refers to continuous variable states that can be fully defined by the first two statistical moments of the bosonic or quadrature field operators \cite{serafini2023quantum, menicucci2011graphical}. 
% This typically simplifies the mathematical representation of the state in comparison to non-Gaussian systems. Often Gaussian states can be fully characterised using finite-dimensional covariance matrices \cite{serafini2023quantum, menicucci2011graphical}. First moments can be arbitrarily tuned through local unitary operations, which do not influence any quantity related to entanglement or mixedness \cite{horodecki2009quantum}.  In general, and throughout this work, the first moments can be set to $0$ without loss of generality. Thus
 First moments can be arbitrarily tuned through local unitary operations, which do not influence any quantity related to entanglement or mixedness \cite{horodecki2009quantum}.  In general, and throughout this work, the first moments can be set to $0$ without loss of generality. Thus
we focus our analysis only with the covariance matrix of the output state.

\begin{figure*}
	\centering
		\begin{subfigure}{\columnwidth}
			\includegraphics[width=\columnwidth]{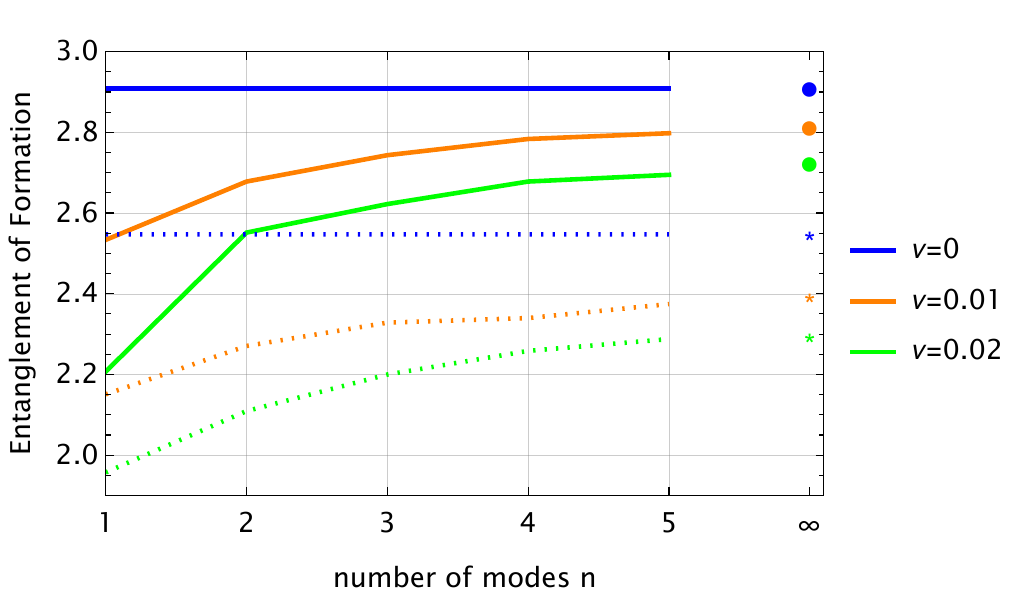}
			\caption{Entanglement of Formation \label{fig:Entanglement}}
		\end{subfigure}
  		\begin{subfigure}{\columnwidth}
			\includegraphics[width=\columnwidth]{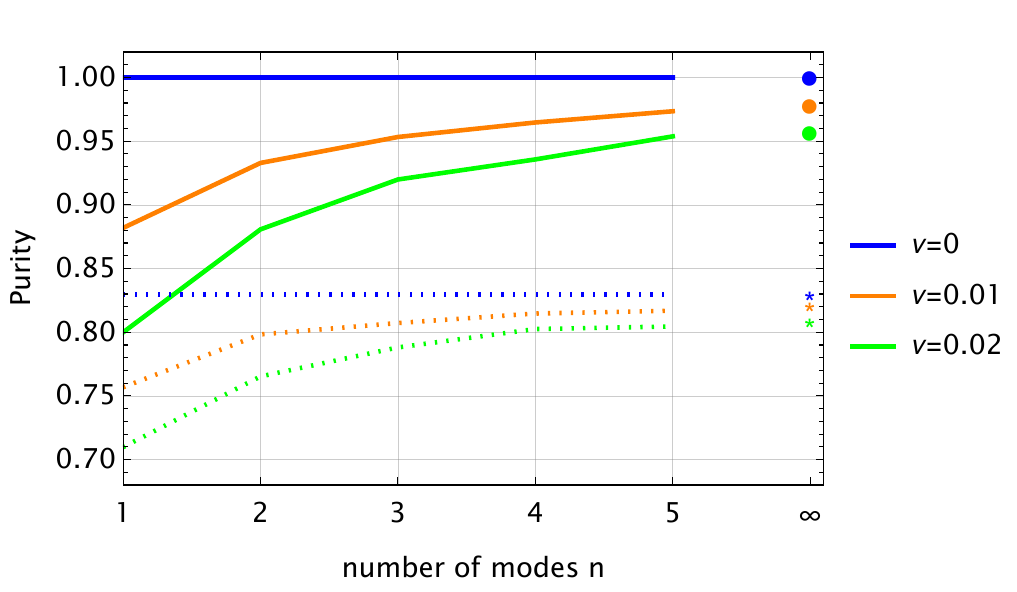}
			\caption{Purity \label{fig:Purity}}
		\end{subfigure}
		\begin{subfigure}{\columnwidth}
			\includegraphics[width=\columnwidth]{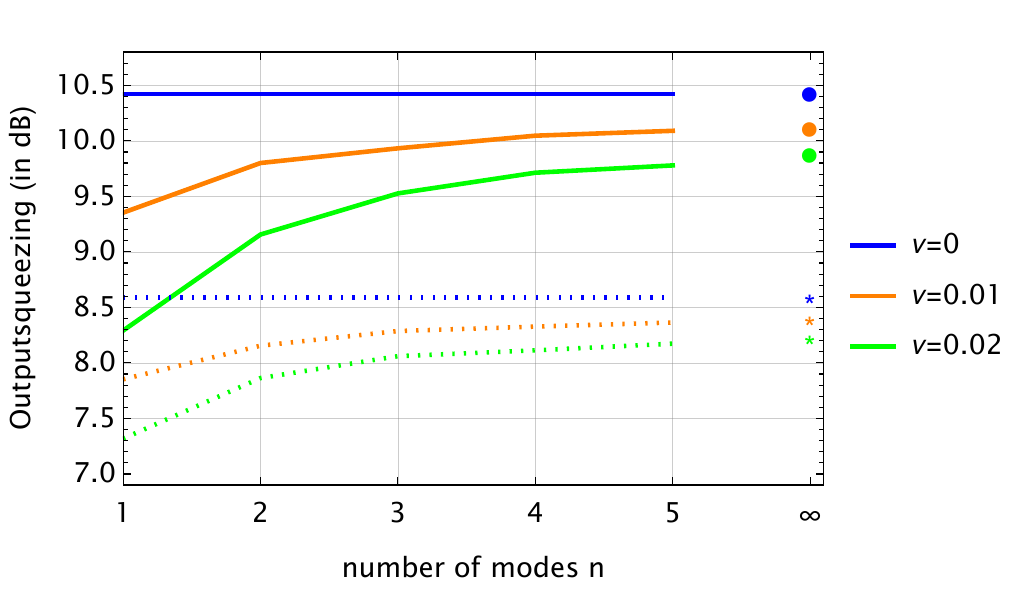}
			\caption{Output Squeezing \label{fig:Output squeezing}}
		\end{subfigure}
		\begin{subfigure}{\columnwidth}
			\includegraphics[width=\columnwidth]{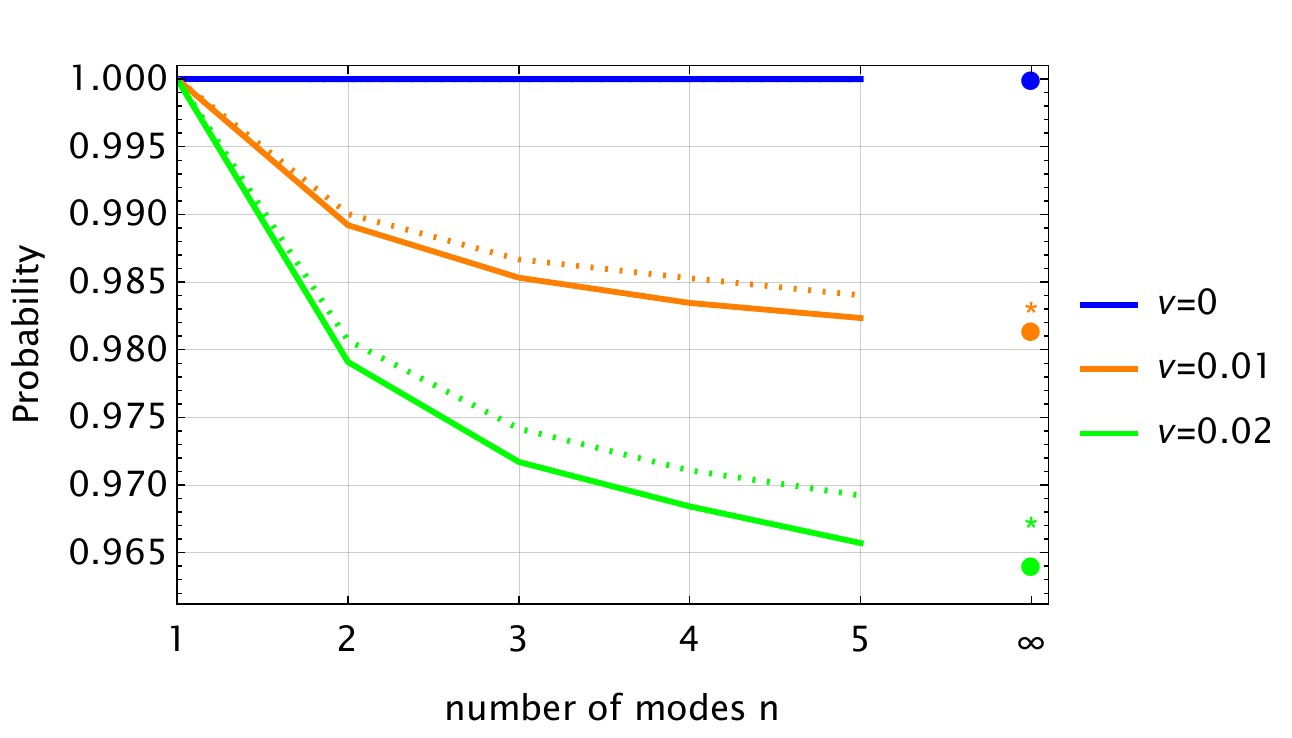}
			\caption{Probability \label{fig:probabilitysuccess}}
		\end{subfigure}
		\caption{Entanglement, purity and output squeezing improvements when applying unitary averaging to the two-mode squeezed vacuum state with  success probability. \ref{fig:Entanglement} EoF is maximum with zero-noise level. As the level of noise increases, our protocol progressively transmits more entanglement with the number of modes. \ref{fig:Purity} Purity is at its peak in a noise-free scenario. As the level of noise increases, our protocol systematically enhances purity with higher number of modes. \ref{fig:Output squeezing} Similarly, output squeezing is same as input squeezing in noise-free region. The effectiveness of output squeezing improves with  higher number of modes with averaging. It is worth noting that the protocol does not lead to squeezing degradation as purity increases. In \ref{fig:Entanglement}, \ref{fig:Purity} \& \ref{fig:Output squeezing}, the results display numerical data both with (dotted lines) and without (solid lines) loss for both practical and asymptotic limit while the input squeezing is $10.4$ dB and loss of $10\%$. \ref{fig:probabilitysuccess}  gives the probability of success when averaging over only few modes and in the asymptotic limit with input squeezing $10.4$ dB. In the supplementary material \cite{supplementary_material}, we showed numerical and analytical results are in perfect agreement.
  \label{fig:Entanglement, purity and squeezing plots}}
		\end{figure*}
   % \ref{fig:probabilitysuccess} Probability of successfully heralding on the vacuum state when applying unitary averaging to the two-mode squeezed vacuum state.
%{\color{blue} Ryan: why not show it for the same 10.4 dB used for all the other plots? Also given all results are function of the variance, we should probably show this as a function of the variance too. }
 %  \begin{figure*}
	% \centering
	% 	\begin{subfigure}{\columnwidth}
	% 		\includegraphics[width=\columnwidth]{figures/pos_1.2_.pdf}
	% 		\caption{\label{fig:Psuccess}}
	% 	\end{subfigure}
	% 	\begin{subfigure}{\columnwidth}
	% 		\includegraphics[width=\columnwidth]{figures/pos_Infinity_with_variance.pdf}
	% 		\caption{\label{fig:Psuccess large n}}
	% 	\end{subfigure}
	% 	\caption{Probability of successfully heralding on the vacuum state when applying unitary averaging to the two-mode squeezed vacuum state.  \ref{fig:Psuccess} gives the probability of success gives the probability of success when averaging over only few modes. The numerical and analytical results are in perfect agreement with input squeezing $10.4$ dB. \ref{fig:Psuccess large n} gives the probability  in the asymptotic limit with input squeezing $10 .4$ dB. \label{fig:Probability of success figures}}
	% \end{figure*}

The output state (\ref{outputeq}) covariance matrix is,
\begin{equation}
     \Sigma_{\text{out}(4 \cross 4)} = \left\langle\begin{pmatrix}
          A  &  C  \\
          C^{T} &  B
     \end{pmatrix}\right\rangle \label{eq:covariance matrix}
 \end{equation}
where,
\begin{align}
    A = B 
    & = \begin{pmatrix}
        \frac{1+ \tanh{r^{'}}^{2}}{1 - \tanh{r^{'}}^{2}} & 0\\
        0 &  \frac{1+ \tanh{r^{'}}^{2}}{1 - \tanh{r^{'}}^{2}}
    \end{pmatrix} \label{aeq}
\end{align}

\begin{align}
    C   
    & = \begin{pmatrix}
       - \frac{2\tanh{r^{'}}}{1 - \tanh{r^{'}}^{2}}\cos\left(\phi_{\beta}\right) & 0\\
0 & \frac{2\tanh{r^{'}}}{1 - \tanh{r^{'}}^{2}}\cos\left(\phi_{\beta}\right) \label{ceq}
    \end{pmatrix}
\end{align}
where $C = C^{T}$. We now employ a small-angle approximations for the noise terms $\phi_{j}$ to approximate $\tanh{r'}$ and $e^{i \phi_{\beta}}$ in order to establish an analytical expression for the unitary averaging model in the low noise limit. If we assume the individual random phases $\phi_{j}$ are independent Gaussian parameters with mean zero and variance $v$, the expectation values within the covariance matrix can be approximated as
\begin{align}
    \left\langle\tanh{r'}\right\rangle \approx& \Big(1 - \big(\frac{v}{2} - \frac{v}{2n}\big)\Big)\tanh{r}, \\
    \left\langle\cos\left(\phi_{\beta}\right)\right\rangle \approx& \cos{\left(\sqrt{\frac{v}{n}}\right)}
\end{align}
where we have assumed $v\ll1$ and
\begin{equation}
\left\langle\frac{2\tanh{r^{'}}}{1 - \tanh{r^{'}}^{2}}\cos\left(\phi_{\beta}\right)\right\rangle \approx \frac{2\left\langle\tanh{r^{'}}\right\rangle}{1 - \left\langle\tanh{r^{'}}\right\rangle^{2}}\left\langle\cos\left(\phi_{\beta}\right)\right\rangle
\end{equation}

% Our analysis will only fully characterise the state if the conditional state is Gaussian which, as discussed earlier is only approximately true. Specifically, in scenarios where all three prerequisites- Gaussian states, Gaussian evolutions and Gaussian measurements, then the conditional state will be Gaussian \cite{niset2009no, weedbrook2012gaussian}. But if one does not have these three pre-conditions, then the output may or may not be Gaussian.  In the case of the vacuum detection, it is a corner case where the conditional state on finding the vacuum happens to be Gaussian.  The conditional state of the failure outcome (i.e. not measuring the vacuum) will not be Gaussian. Here, we are exclusively focused on success modes, which entails our consideration of Gaussian outcomes from our model.

We compute these analytic approximations and the numerical simulations of the exact covariance matrix (given by Equations \ref{eq:covariance matrix}-\ref{ceq}) and analyse the success probability, purity, output squeezing, and the entanglement both with and without applying UA. This provides insights into how our protocol protects the channel against stochastic phase noise. Numerical simulations are performed using \textit{Mathematica}, where multiple samples of covariance matrices are considered, each subject to Gaussian, independent, stochastic phase noise with mean zero and variance $v$.

%\section{Purity, squeezing and entanglement}
\textit{Purity, squeezing and entanglement.-}
\begin{table}
  \begin{tabular}{ p{2cm} p{2cm} p{2cm} p{2cm}  }
 \hline
  \hline
 Squeezing factor(r) & Input squeezing & Output squeezing ($n=1$) & Output squeezing ($n=5$)\\
 \hline
0.5 & 4.34   & 4.09 &   4.27\\
 1 &  8.69  & 6.84 & 8.17 \\
 1.2 & 10.43 & 6.87 &  9.39\\
 1.5   & 12.03 &  6.16 &  10.43\\
2 &   17.38  & 2.49 & 9.03\\
\hline
\hline
\end{tabular}
  \caption{ The squeezing here is reported in decibels. Comparing input to output squeezing. The `r' value determines the input squeezing. The $n=1$ case gives the original channel output squeezing (without averaging), and the $n=5$ case shows the improvement with modest correction (with UA) while the variance is $0.01$.}
    \label{table}
\end{table}

In realistic systems, phase fluctuations will act to diminish the achieved squeezing and make the state no longer minimum uncertainty \cite{walls1983squeezed}.
%Presently available squeezed states often exhibit significant impurity, primarily arising from the effects of decoherence and dissipation during their preparation from sources. 
Our protocol has the ability to address the issues by effectively lowering the phase noise leading to simultaneously improving purity, squeezing and entanglement through a noisy channel. We are modelling this with the stochastic phases which vary from shot-to-shot. 

%improving purity while minimise anti-squeezing effects and minimally impacting squeezing through a noisy channel. 

Measuring the variances of the output state, given its approximate Gaussian nature, serves as a comprehensive means to assess both output squeezing and purity \cite{weedbrook2012gaussian, ferraro2005gaussian}. For a Gaussian state the product of uncertainties is directly related to the purity in the resultant state. Table \ref{table} clearly illustrates the significant enhancement in squeezing achieved through the channel when utilising the passive averaging protocol. Of particular note is that the maximum possible output squeezing without averaging ($n =1$) is significantly lower than the maximum output squeezing that can be achieved with averaging ($n=5$).

Quantifying entanglement is a non-trivial task as various measures have been defined with distinct operational meanings \cite{vedral1997quantifying, horodecki2009quantum, bowen2004experimental, tserkis2017quantifying, tserkis2019quantifying}. Here we use the entanglement of formation (EoF), which is valued for its important physical significance in quantum technologies \cite{chitambar2019quantum} because it aligns with the entanglement cost \cite{hayden2001asymptotic}. 
% The entanglement cost is a measure of the minimum entanglement required (representing the cost of quantum resources) to create that state.
We utilise the general formula detailed in Reference \cite{tserkis2017quantifying} to quantify the level of entanglement present in our output state. 
The EoF, purity and squeezing are plotted with number of modes $n$ in Fig.\ref{fig:Entanglement, purity and squeezing plots}. This shows the practical advantages of the model with significant improvements in the output squeezing, purity and EoF, even for modest $n$. The output squeezing is seen to increase by $1.5$ to $2$ dB even for a single extra channel ($n=2$). Further improvement is seen with larger n. 
Given that purity and entanglement are interlinked, the channel yields a significant enhancement in entanglement.

% The solid and dashed lines, representing numerical and analytical results respectively, are an agreement in the small noise (small $v$) limit we consider here. 

% The squeezing will be seen in the variances of the difference quadratures, which can be visualise from the covariance matrix elements. Here, we present an analytical formula for output squeezing,
We have also calculated the probability of successfully heralding on the vacuum state necessary to produce the lower noise, higher entangled state. The resulting probability scaling is shown in Figure \ref{fig:probabilitysuccess}. These show that the probability cost to yield the significant improvements in entanglement, purity and squeezing from the channel are only modest, with at worst a $96.5\%$ chance of successfully heralding on the desired vacuum state, low noise limit we are primarily concerned with here. Importantly, the probability remains at a useful level even in the large $n$ limit and large noise, whilst maintaining non-zero entanglement as shown in the supplementary material \cite{supplementary_material}.

\textit{ Loss errors.-} Loss within the channel and vacuum detectors is modelled as beamsplitters before each elements.  These beamsplitters mix the channel with the vacuum at a rate  $\sqrt{\gamma}$ giving a loss probability $\gamma$ \cite{albert2018performance, albert2022bosonic}. This operation, however, commutes through the entire transformation applied to the input mode until it is the first operation acting on the input state, provided we consider the loss to act on all modes equally, including the final conditional output state. Hence, any loss acting on the channel will be equivalent to loss within the initial state. Note that as most inputs are in the vacuum it only looks like loss acting on the input two mode squeezed state. This tells us that loss in the vacuum projection detectors and in the initial state do not impact the ability to correct the phase noise.

 \textit{Conclusion.-} Our protocol can be effectively implemented within state-of-the-art quantum communication and computing platforms, as it relies on readily available Gaussian states, linear optics setups, and vacuum projection, all commonly found in various practical bosonic systems.

In summary, we have presented a passive unitary averaging scheme with vacuum detection acting on one arm of a two-mode squeezed state. We found it successfully limits the effect of phase noise within the channel. Over an order of magnitude of noise reduction was achieved surprisingly with an increase in squeezing as well as purity, and enhancing levels of entanglement available in realistic systems. We expect that the UA techniques described here will find immediate applications in quantum communications and optical quantum computing schemes.

\textit{Acknowledgement.-} We thank Deepesh Singh for helpful discussion about unitary averaging.
SNS  expresses gratitude to Giacomo Pantaleoni, Fumiya Hanamura and Takaya Matsuura for their  discussions on Gaussian states. SNS was supported by the Sydney Quantum Academy, Sydney, NSW, Australia.  
This work was partially supported by the Australian Research Council Centre of Excellence for Quantum Computation and Communication Technology (Project No.CE110001027).
\bibliography{mainv2}

\clearpage

\onecolumngrid

\renewcommand{\bibnumfmt}[1]{[S#1]}{}\renewcommand{\citenumfont}[1]{S#1}
%\section{Appendices}
\clearpage

\begin{center}
{\Large{Supplemental Material for ``Improving Continuous-variable Quantum Channels with Unitary Averaging"}}
\end{center}
\section{Passive unitary averaging with loss on CV systems}
In the main text \cite{maindraft}, we operate under the assumption of zero channel loss within the model. We clarify in the main text \cite{maindraft} that loss occurring within the channel or detectors is equivalent to loss within the initial state. As most inputs are in the vacuum, it looks like loss acting only on the two-mode squeezed vacuum state. This equivalence is true provided there is the same amount of end-to-end loss, $1 - \eta$, on all modes. If this is true then the output when loss is distributed through the circuit is equivalent to the output when the same loss is applied to the input state, which is then propagated through a lossless circuit.
In this section, we calculate the output for the two mode squeezed vacuum state which undergoes loss, and then passes through the passive, linear optical unitary averaging scheme. For the sake of mathematical simplicity, we consider the case where the mode $\hat{a}_{1}$ undergoes loss. The calculation of the covariance matrix is easier when approached through Heisenberg picture. The resulting channel transmissivity is $\eta$.

 In our case, the two-mode squeezer acts on the modes labeled $\hat{a}_{0}$ and $\hat{a}_{1}$. Then, the transmitted mode  becomes,
 \begin{align}
 \hat{a}'_{1} &= \sqrt{\eta} \hat{a}_{1} + \sqrt{1-\eta}v_{1} \label{lossmodes}
\end{align}
where $v_{1}$ is vacuum noise of mode $\hat{a}_{1}$.
After passing through the loss, interferometer and heralding on the last $(n-1)$ modes, the output state is
 \begin{align}
    \ket{\psi_{\text{out}}}=\left(\otimes_{j=2}^n\bra{0}_j\right)\hat{L}_{1}\hat{U} \hat{S}(r)\ket{0}\ket{0} \otimes \ket{0}^{\otimes(n-1)}
 \end{align}
where $\hat{L}_{1}$ is loss acting on mode $\hat{a}_{1}$, $\hat{U} = (\hat{H}^{\dagger}*\hat{R}(\theta)*\hat{H})_{n \times n}$,  $\hat{H}$ is the Hadamard transforms acting as the encoding and decoding circuits, and $\hat{R}(\theta)$ implements stochastic phase transformations on each mode, representing the noise.

The output state is,
 \begin{widetext}
     \begin{equation}
     \ket{\psi_{\text{out}}} = \frac{1}{\mathbf{N}}(\cosh{r})^{-1} \sum^{\infty}_{N=0} \sum^{N}_{K=0}  \frac{\sqrt{N!}}{\sqrt{K!}\sqrt{(N-K)!}} (-1)^{N} (\tanh{r})^{N} (\alpha e^{i\phi_{\beta}})^{N-K} \eta^{\frac{N-K}{2}} (1 - \eta)^{\frac{K}{2}} \ket{N, N-K, K} \label{outputlosseq}
 \end{equation}

\end{widetext}
where $\mathbf{N}$ is a normalisation constant related to probability of success, $K$ is the lossy mode and $\alpha e^{i\phi_{\beta}}=\frac{e^{ i \phi_{1}} + e^{ i \phi_{2}} +...+e^{ i \phi_{n}}}{n}$.

The normalisation constant for the output state $(\ref{outputlosseq})$ is,
\begin{align}
    \mathbf{N} 
    & = \sqrt{\frac{(\sech{r})^{2}}{-1 + (1 + (-1 + |\alpha e^{i\phi_{\beta}}|^{2} )\eta)(\tanh{r})^{2}}}
\end{align}
We again use the covariance matrix formalism \cite{weedbrook2012gaussian,serafini2023quantum} to characterise the output state after introducing loss into the input two-mode squeezed vacuum state. We present the numerical results with loss in the main text \cite{maindraft}.

 \section{Several plots showing numerical and analytical agreement}
\begin{figure*}
	\centering
  \begin{subfigure}{\columnwidth}
			\includegraphics[width=\columnwidth]{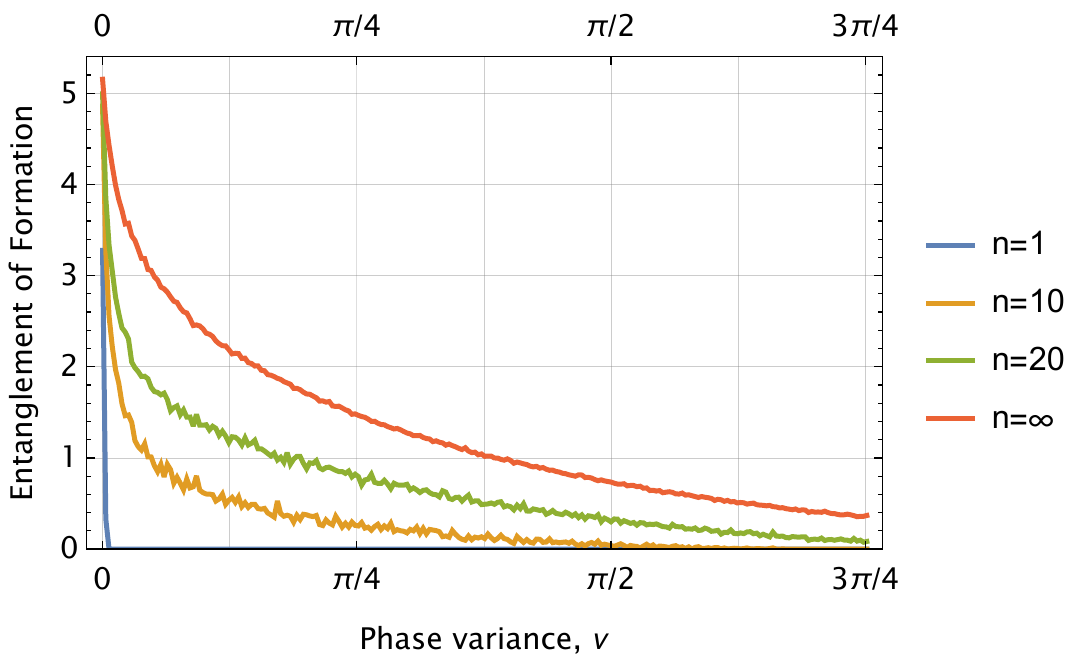}
			\caption{Entanglement of Formation in the $n\rightarrow\infty$ limit \label{fig:Entanglement large n}}
		\end{subfigure}
  \begin{subfigure}{\columnwidth}
			\includegraphics[width=\columnwidth]{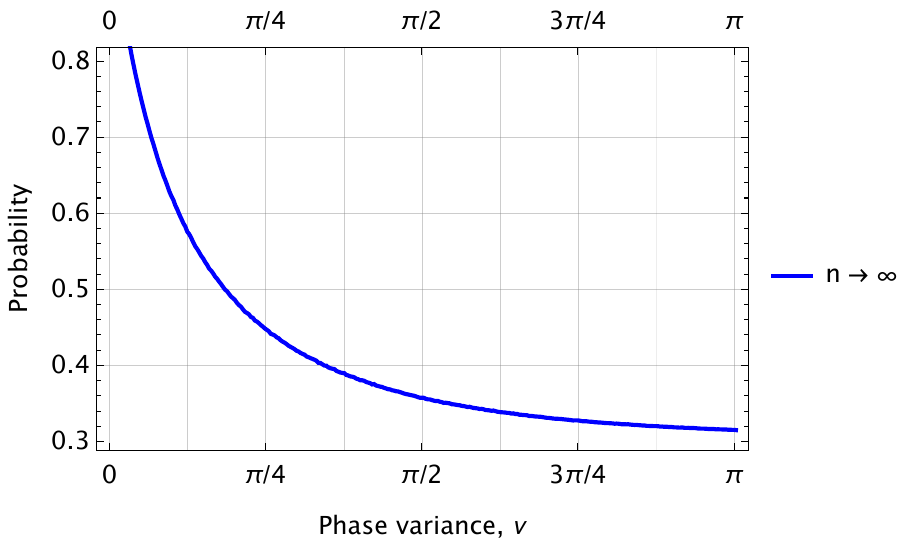}
			\caption{Probability of Success in the $n\rightarrow\infty$ limit\label{fig:Psuccess large n}}
		\end{subfigure}
		\caption{Entanglement and probability in the asymptotic limit  with input squeezing $10.4$ dB.   \ref{fig:Entanglement large n} Entanglement is enhancing with increased phase variance $n\rightarrow\infty$ limit.  \ref{fig:Psuccess large n} Probability is non-zero with sufficiently high noise in the asymptotic n limit. \label{fig:Probability of success figures}}
	\end{figure*}

In this section, we present various plots demonstrating the alignment between analytical and numerical results, specifically fitted to showcase advancements in practical technologies. Additionally, we demonstrate the asymptotic nature of our protocol. In Fig.\ref{fig:Entanglement, purity and squeezing plots}, we show numerical and analytical results are in perfect agreement in the low noise regime.
As the noise level increases, our protocol dynamically increases the transmission of entanglement, purity, and squeezing compared to a noisy channel, with at worst 93\% success probability for the noise levels considered, as shown in Figures \ref{fig:Entanglement, purity and squeezing plots} and \ref{fig:Probability of success figures}. Fig.\ref{fig:Entanglement, purity and squeezing and prob with r plots} shows how entanglement, purity, output squeezing and success probability change with varying levels of input squeezing with low noise. Entanglement and squeezing initially increase with input squeezing but diminish at higher input squeezing levels, although it still improves with an increased number of modes. Purity decreases as squeezing increases but again improved with a larger number of modes. 
In Fig.\ref{fig:Entanglement, purity and squeezing and prob with r in the limit of high n plots} we illustrate the robust aspect of our protocol, which remains benifitial even under high squeezing and demonstrate its behaviour when utilising a large number of modes.

\begin{figure*}
	\centering
		\begin{subfigure}{\columnwidth}
			\includegraphics[width=\columnwidth]{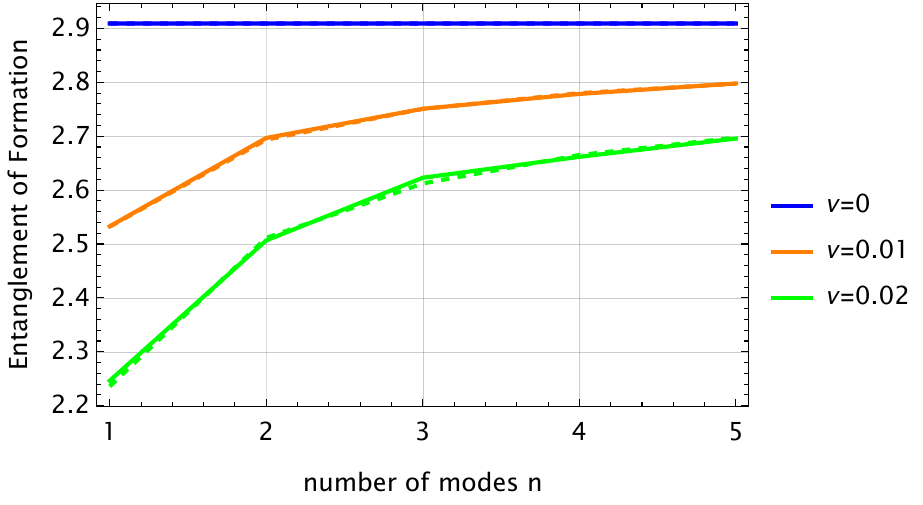}
			\caption{Entanglement of Formation \label{fig:Entanglement}}
		\end{subfigure}
  		\begin{subfigure}{\columnwidth}
			\includegraphics[width=\columnwidth]{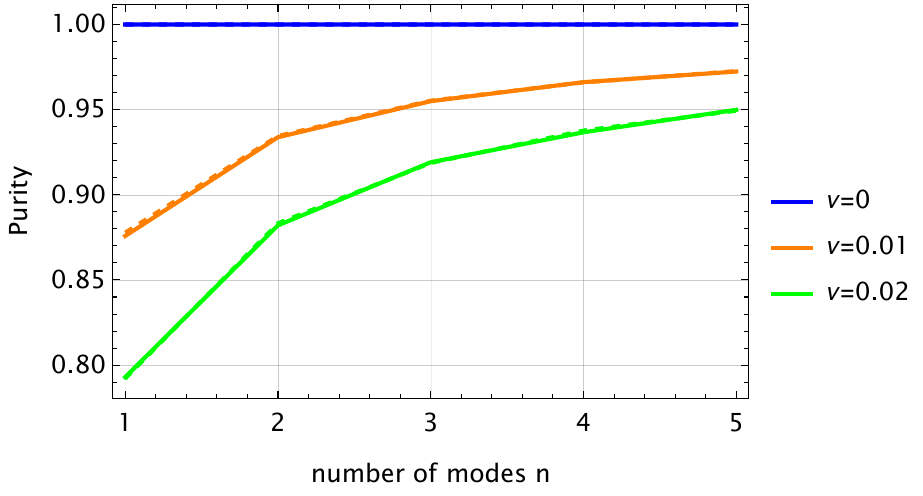}
			\caption{Purity \label{fig:Purity}}
		\end{subfigure}
		\begin{subfigure}{\columnwidth}
			\includegraphics[width=\columnwidth]{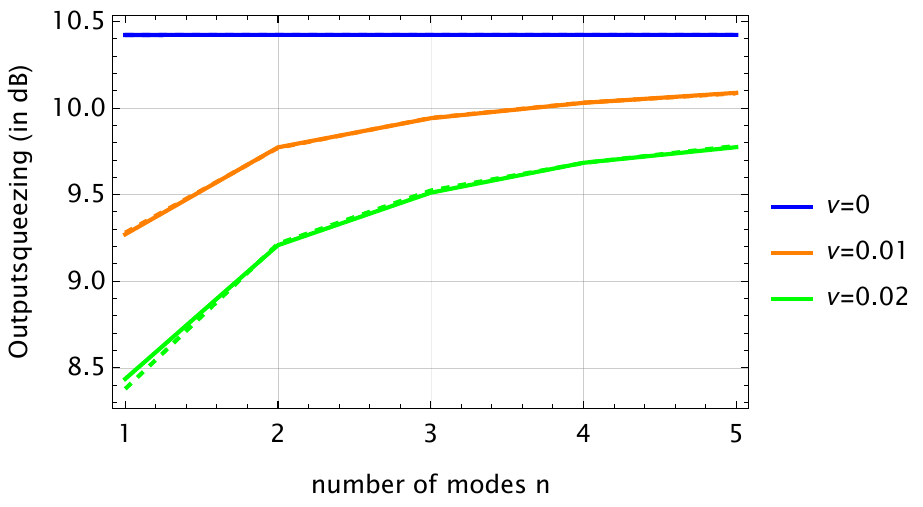}
			\caption{Output Squeezing \label{fig:Output squeezing}}
		\end{subfigure}
		\begin{subfigure}{\columnwidth}
			\includegraphics[width=\columnwidth]{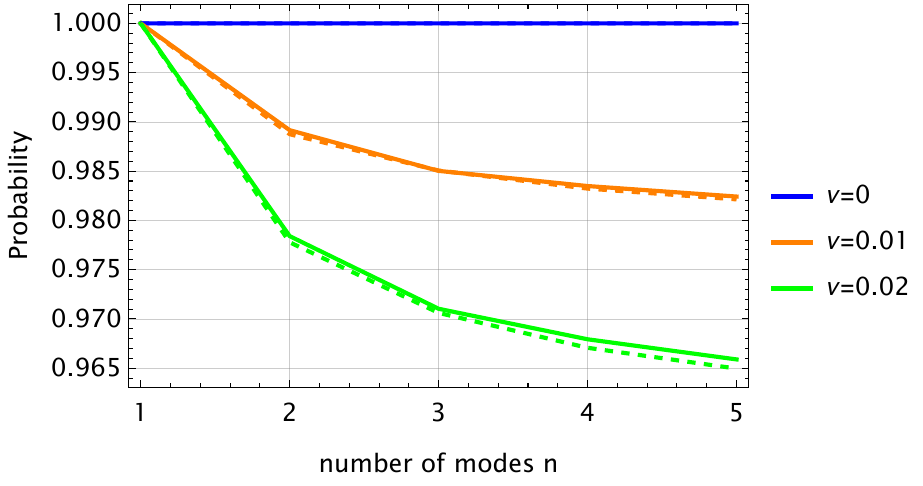}
			\caption{Probability of Success\label{fig:Psuccess}}
		\end{subfigure}
		\caption{Entanglement, purity and output squeezing improvements with success probability when applying unitary averaging to the two-mode squeezed vacuum state. \ref{fig:Entanglement} EoF is maximum with zero-noise level. As the level of noise increases, our protocol progressively transmits more entanglement with the number of modes. \ref{fig:Purity} Purity is at its peak in a noise-free scenario. As the level of noise increases, our protocol systematically enhances purity with higher number of modes. \ref{fig:Output squeezing} Similarly, output squeezing is same as input squeezing in noise-free region. The effectiveness of output squeezing improves with  higher number of modes with averaging. It is worth noting that the protocol does not lead to squeezing degradation as purity increases. In \ref{fig:Entanglement}, \ref{fig:Purity} \& \ref{fig:Output squeezing}, the numerical (solid lines) and analytical (dotted lines) results are in perfect agreement with input squeezing $10.4$ dB. \ref{fig:Psuccess} gives the probability of success when averaging over only few modes. 
  The numerical (solid lines) and analytical (dotted lines) results are in perfect agreement with input squeezing $10.4$ dB.
  \label{fig:Entanglement, purity and squeezing plots}}
\end{figure*}

\begin{figure*}
	\centering
		\begin{subfigure}{\columnwidth}
			\includegraphics[width=\columnwidth]{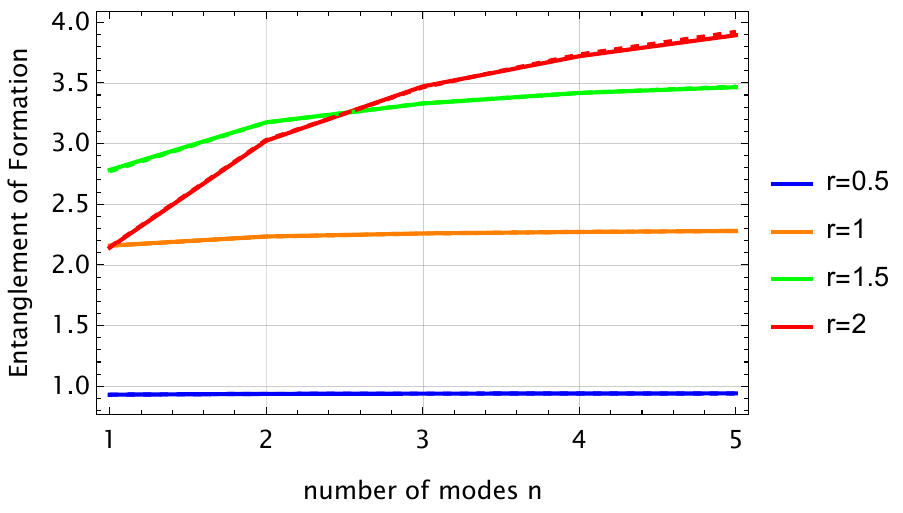}
			\caption{Entanglement of Formation with different ``r". \label{fig:Entanglement with r}}
		\end{subfigure}
  		\begin{subfigure}{\columnwidth}
			\includegraphics[width=\columnwidth]{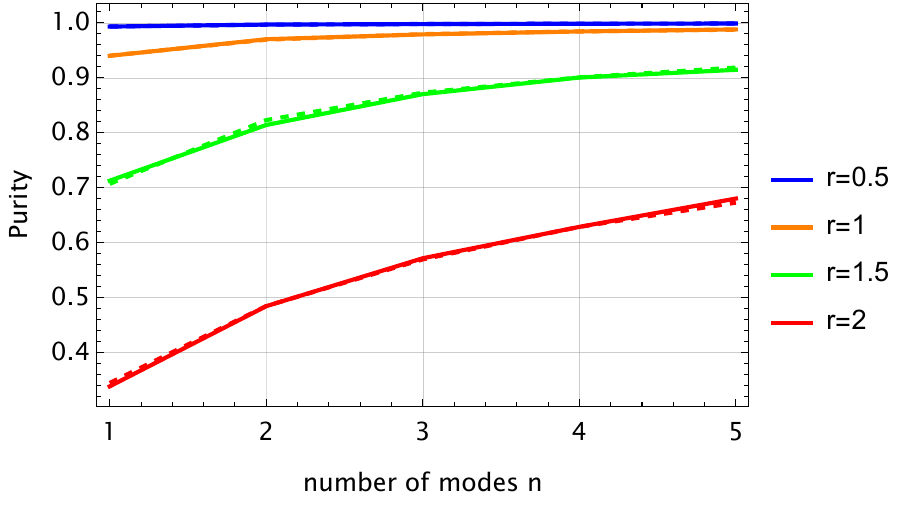}
			\caption{Purity with different ``r".\label{fig:Purity with r}}
		\end{subfigure}
		\begin{subfigure}{\columnwidth}
			\includegraphics[width=\columnwidth]{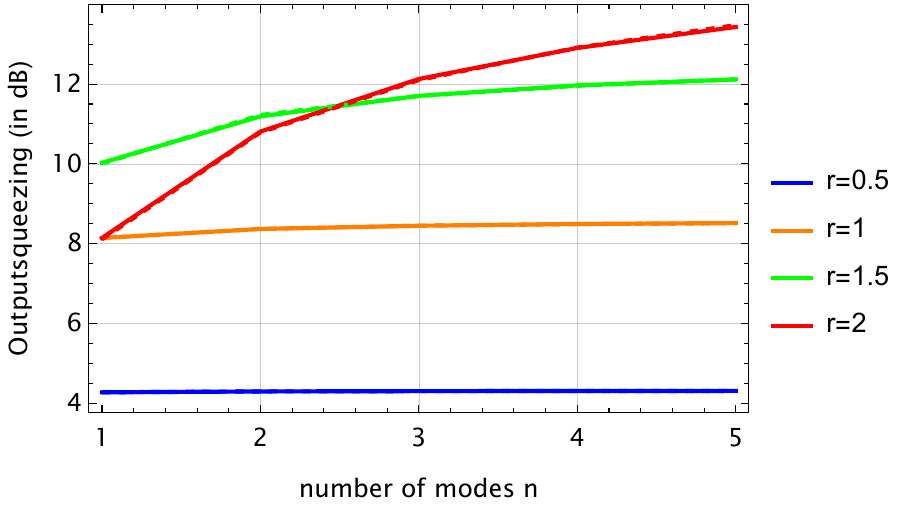}
			\caption{Output Squeezing with different ``r". \label{fig:Output squeezing with r}}
		\end{subfigure}
		\begin{subfigure}{\columnwidth}
			\includegraphics[width=\columnwidth]{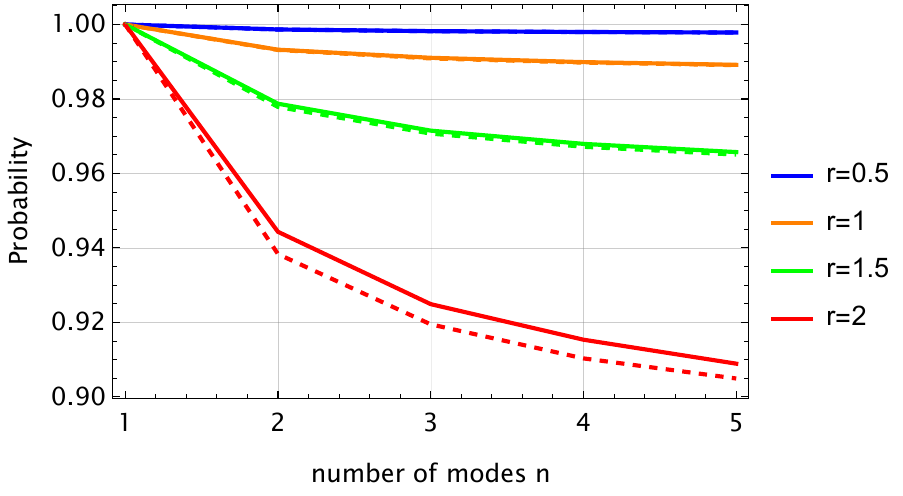}
			\caption{Probability of success with different ``r". \label{fig:prob with r}}
		\end{subfigure}
		\caption{Entanglement, purity and output squeezing varies alongside the success probability when applying unitary averaging to the two-mode squeezed vacuum state with varying levels of input squeezing in the low noise limit. \ref{fig:Entanglement with r} EoF began to decline at input squeezing $r =2$. \ref{fig:Purity with r} Purity decays with the increasing input squeezing and interestingly even at the high level of input squeezing, the output state is not completely mixed. \ref{fig:Output squeezing with r} Similarly, output squeezing started decreasing after reaching input squeezing  $r =1.5$. In \ref{fig:Entanglement with r}, \ref{fig:Purity with r} \& \ref{fig:Output squeezing with r}, the numerical (solid lines) and analytical (dotted lines) results are in perfect agreement with noise $0.01$. \ref{fig:prob with r} Numerical Probability looses agreement with analytical probability from $r =2$ with noise $0.01$. 
  \label{fig:Entanglement, purity and squeezing and prob with r plots}}
\end{figure*}

\begin{figure*}
	\centering
		\begin{subfigure}{\columnwidth}
			\includegraphics[width=\columnwidth]{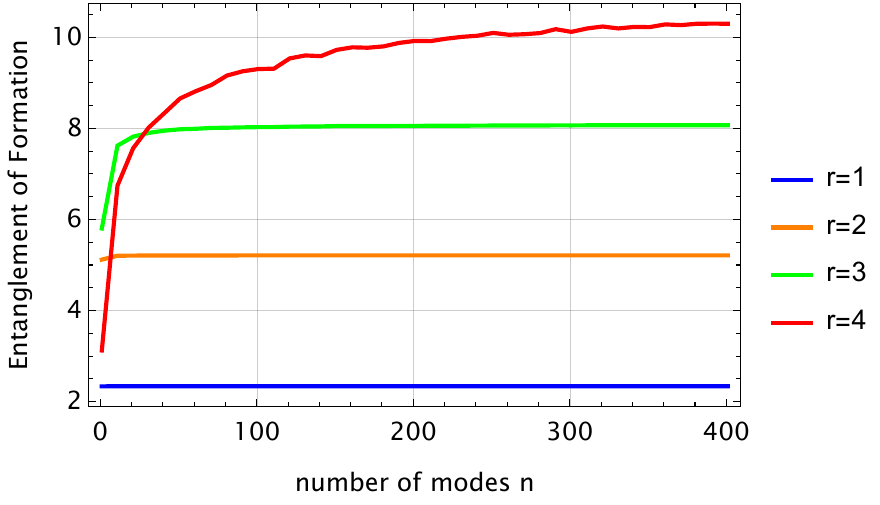}
			\caption{Entanglement of Formation with different ``r". \label{fig:Entanglement with r in the limit of high n}}
		\end{subfigure}
  		\begin{subfigure}{\columnwidth}
			\includegraphics[width=\columnwidth]{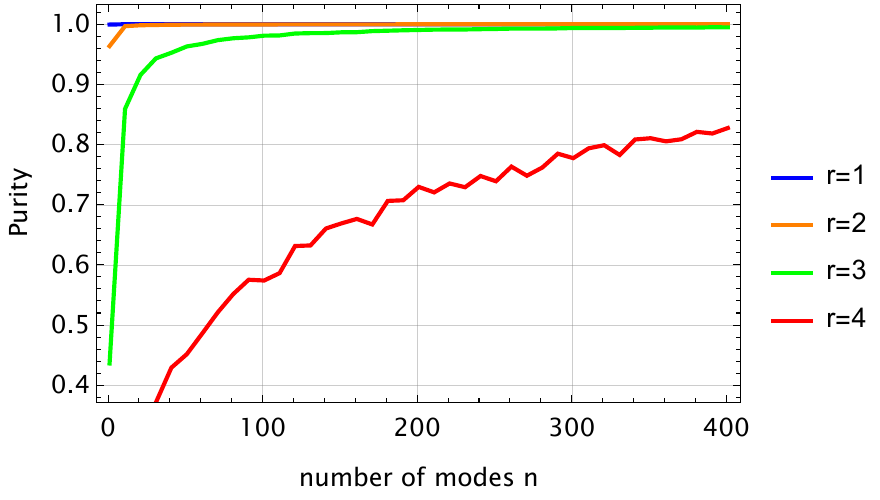}
			\caption{Purity with different ``r".\label{fig:Purity with r in the limit of high n}}
		\end{subfigure}
		\begin{subfigure}{\columnwidth}
			\includegraphics[width=\columnwidth]{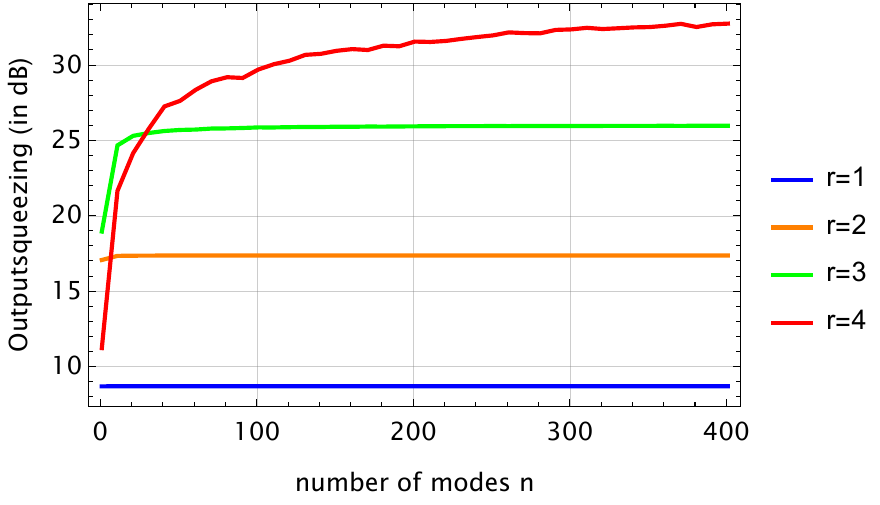}
			\caption{Output Squeezing with different ``r". \label{fig:Output squeezing with r in the limit of high n}}
		\end{subfigure}
		\begin{subfigure}{\columnwidth}
			\includegraphics[width=\columnwidth]{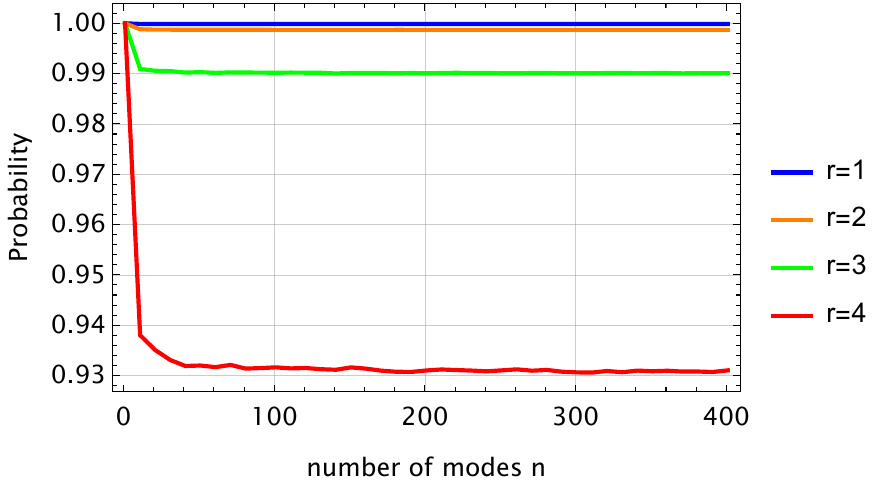}
			\caption{Probability of Success with different ``r". \label{fig:prob with r in the limit of high n}}
		\end{subfigure}
    		\caption{Entanglement, purity and output squeezing varies alongside the success probability when applying unitary averaging to the two-mode squeezed vacuum state with varying levels of input squeezing in the asymptotic limit with noise $0.01$. \ref{fig:Entanglement with r in the limit of high n} \& \ref{fig:Output squeezing with r in the limit of high n} EoF and output squeezing behave similarly with the higher number of modes with high input squeezing. \ref{fig:Purity with r in the limit of high n} Purity increases with the number of modes and getting saturated with the higher number of modes.  \ref{fig:prob with r in the limit of high n} Probability of success drops massively for input squeezing $r=4$, and getting saturated with higher $n$.
  \label{fig:Entanglement, purity and squeezing and prob with r in the limit of high n plots}}
\end{figure*}

\end{document}